\documentstyle[twocolumn,prl,aps]{revtex} 
 
\begin{document} 
\twocolumn[\hsize\textwidth\columnwidth\hsize\csname @twocolumnfalse\endcsname
\draft 
\preprint{} 
\title{Reflection Symmetry and Quantized Hall Resistivity near Quantum Hall
Transition}
\author{D.N. Sheng and Z. Y. Weng} 
\address{ Texas Center for Superconductivity, 
University of Houston, Houston, TX 77204-5506$^*$ \\
and National Center for Theoretical Science, P. O. Box 2-131, Hsinchu, Taiwan 300, R.O.C. }  
\maketitle 
\date{today}
\begin{abstract} 
We present a direct numerical evidence for reflection symmetry of 
longitudinal resistivity $\rho_{xx}$ and quantized Hall resistivity $\rho_{xy}$ 
near the transition between $\nu=1$ quantum Hall state and insulator, in accord 
with the recent experiments. Our results show that a universal scaling behavior 
of conductances, $\sigma_{xx}$ and $\sigma_{xy}$, in the transition regime 
decide the reflection symmetry of $\rho_{xx}$ and quantization of $\rho_{xy}$, 
independent of particle-hole symmetry. We also find that in insulating phase
away from the transition region $\rho_{xy}$ deviates from the quantization and
diverges with $\rho_{xx}$. 

\end{abstract}
]

A reflection symmetry has been recently revealed \cite{shar1,shar2} in the 
transition between the quantum Hall (QH) states and the 
high-magnetic-field insulator: e.g., the longitudinal resistivity in the
insulator and its inverse in the neighboring $\nu=1$ QH state satisfies a 
relation 
\begin{equation}\label{1}
\rho_{xx}(\Delta n_{\nu})=1/\rho_{xx}(-\Delta 
n_{\nu})
\end{equation}
where $\Delta n_{\nu}$ is the Landau level (LL) filling number $n_{\nu}$ 
measured from the critical point $n_{\nu c}$. At the same time, the Hall
resistivity remains well quantized, i.e., 
\begin{equation}\label{2}
\rho_{xy}=h/e^2
\end{equation}
over the whole transition regime. Similar symmetry also holds at 
$\nu=1/3\rightarrow 0$ transition\cite{shar1}.
Theoretically, such a reflection symmetry has been conjectured as due to an
underlying charge-flux duality\cite {phs} in the composite boson 
description\cite{klz,prya1} of the QH transitions, which is equivalent to a 
particle-hole symmetry 
in the composite fermion description\cite{jain}. Indeed, at $\nu=1\rightarrow 0$ 
transition, one can easily understand (\ref{1}) if a particle-hole symmetry in 
the fermion description is
assumed in the lowest Landau level (LLL), and with using (\ref{2}). But in a
general situation, disorders do not necessarily retain the particle-hole 
symmetry of the LLL in a microscopic Hamiltonian of 
non-interacting fermions in the presence of magnetic field. It is not 
clear theoretically whether the reflection symmetry can still remain there, 
even though the experiments\cite{shar1,shar2} have clearly shown that critical 
point floats away from the LLL center without affecting the reflection symmetry. 
 
The quantization of Hall resistivity (\ref{2}) beyond the QH plateau regime is
also nontrivial. A quantized Hall insulator has been
discussed based on a phase-incoherent network model\cite{shim,prya2,ruz}, and
it has been also found\cite{prya2} that $\rho_{xy}$ may become divergent if the
quantum interference is taken into account. Whether the quantization of
$\rho_{xy}$ in the transition region is a classical behavior due to the absence 
of quantum interference at a finite temperature as discussed in those approaches
or it is a universal behavior related to the property of quantum critical 
point\cite{prus,kram,bhatt} of the QH state to insulator transition needs an
independent clarification.

In this Letter, we present a direct numerical evidence for the reflection
symmetry (\ref{1}) in the $\nu=1\rightarrow 0$ transition. A quantized 
Hall resistivity in the transition
region is also obtained in accord with (\ref{2}). In particular, all these 
results persist even when disorders are strong enough such that Landau 
level (LL) mixing becomes important. It suggests that the underlying 
particle-hole symmetry in the LLL is not crucial to the reflection symmetry as 
well as the quantization 
of the Hall resistance. These properties are found to be related to a 
universal scaling of conductances near the transition. 
The scaling functions of $\sigma_{xx}^2$ and $\sigma_{xy}*(1-\sigma_{xy})$
(in units of $e^2/h$) are found to be equal to each
other as a universal curve which is symmetric with 
regard to the critical point, independent of sample sizes and disorder 
strengths. Furthermore, beyond the transition regime we find that $\rho_{xy}$
starts to deviate from the quantized value and eventually diverges as $\rho_{xx}\rightarrow \infty$ in insulating regime. 
To test the robust of the reflection symmetry, we also study the
case where the higher QH plateaus ($\nu>1$) are already destroyed at strong 
disorders, with only $\nu=1$ QH plateau left. This situation has been realized 
in a tight-binding lattice model\cite{xie,dns,pbs} and there one finds another
$\nu=1\rightarrow 0$ transition on the high filling number side\cite{dns}. 
While the reflection symmetry and the Hall resistance quantization are still 
robust near the $\nu=1\rightarrow 0$ transition on the low-filling-number side, 
they {\it disappear} at the transition on the high-filling side, suggesting
that the existence of the QH plateau(s) {\it above} the Fermi energy may be
crucial for both the reflection symmetry and the quantization of the Hall 
resistance.

We use a tight-binding Hamiltonian of non-interacting electrons:  
\begin{eqnarray*} 
H=-\sum_{<ij> } e^{i a_{ij}}c_i^+c_j + H.c. +\sum _i w_i c^+_i c_i , \nonumber
\end{eqnarray*} 
where the hopping integral is taken as the unit, and $c_i^+$ is a  
fermionic creation 
operator with $<ij>$  referring to two  nearest neighboring sites.  A uniform  
magnetic flux per plaquette is  given as  $\phi=\sum _ {\Box} a_{ij}=2\pi/M$,
where the summation runs over four links around a plaquette. In the following
we mainly focus on $M=8$ case, while weaker fields with larger $M$ are also 
checked. $w_i$ is a random potential with strength $|w_i|\leq W/2$, and the
white  noise limit is considered with no correlations among different sites
for $w_i$.  In the weak disorder limit, the mixing between Landau levels 
can be neglected so the disorder effect in the LLL still approximately respects
the particle-hole symmetry. With the increase of $W$, Landau levels start to
mix together such that the definition of ``holes'' is no longer meaningful in 
the LLL and the particle-hole symmetry is then removed. In the following, we
mainly study $W=1$ and $W=4$ which represent these two limits. We also focus on 
the $\nu=1\rightarrow 0$ transition which happens at the step of 
the Hall conductance $\sigma_{xy}$ between $e^2/h$ and $0$. $\sigma_{xx}$ at the
critical point scales onto a constant value $\sigma_{xx}=(0.50 \pm 0.02)e^2/h$ 
independent of disorder or magnetic field strength in agreement with earlier 
work\cite{kram,bhatt}.  Here  $\sigma_{xx}$ is calculated using Landauer formula\cite{land}
for square sample size $L\times L$ with leads and averaged over random 
disorder configurations to obtain a required statistical error
(less than $2\%$). It has been checked by us that $\sigma_{xx}$ calculated
in this way agrees with the conductance calculated from Thouless
number\cite{bhatt1} as long as they are scaled to the same value at the critical 
point. Hall conductance $\sigma_{xy}$ is calculated by using Kubo formula (at 
least 2,000 configurations are taken for the largest sample sizes). 

The quantization of the Hall resistivity is clearly shown in Figs. 1a and 1b
near the $\nu=1\rightarrow 0$ transition at $M=8$. Here 
$\rho_{xy}$ is plotted as a function of $\sigma_{xy}$ at 
different sample widths: $L=8,16,24$ and $32$, with $W=1$ and $W=4$, 
respectively. In such a transition regime, $\sigma_{xy} $ changes from
$1$ to $0$ (in units of $e^2/h$), while $\rho_{xy}$ stays almost constant in the 
transition region at the quantized value $1$. Such a quantization remains in spite of
the change of the particle-hole symmetry from $W=1$ to $W=4$: the density of 
states, $D_L(\epsilon)$, is shown in the inserts of Figs. 1a and 1b, and in 
contrast to $W=1$ where the LLL is well 
separated from higher LL's by a gap, two lowest LLs are mixed at $W=4$
where the particle-hole symmetry is obviously absent.

The reflection symmetry of the longitudinal resistivity is  
shown in Fig. 2 at $W=1$ and $W=4$, respectively. 
Here the resistivity is plotted as a function of $\Delta \epsilon$, i.e., the
Fermi energy $\epsilon$ measured from the critical point $\epsilon_c$.
As shown in Fig. 2, $\rho_{xx}(\Delta \epsilon )$ and $1/\rho_{xx} 
(-\Delta\epsilon) $ are right on top of each other, demonstrating the 
reflection symmetry of $\rho_{xx}$ at a sample width $L=32$.
In both cases, the reflection symmetry is exhibited over a very broad
region: $\rho _{xx} $ changes more than one order of the magnitude in
the insulating region similar to the experimental 
observation\cite{shar1,shar2} (the region with reflection symmetry
generally grows with the sample size $L$ in our calculation).

The quantization of the Hall resistance implies a semi-circle law: 
\begin{equation}\label{3}
\sigma_{xx}^2 +(\sigma_{xy}-\frac 1 2)^2=\frac 1 4
\end{equation}
or $\sigma_{xx}^2=\sigma_{xy}*(1-\sigma_{xy})$, which was previously obtained 
based on a semiclassical treatment of the QH edge states\cite {ruz}. 
To make sure if this effect is an intrinsic property of a quantum critical point
instead of a finite-size effect, we have checked the
scaling behavior of both $\sigma_{xx}$ and $\sigma_{xy}$.
It has been well established\cite {kram} that $\nu=1\rightarrow 0$ transition 
should satisfy a one-parameter scaling law:
both $\sigma_{xx}$ and $\sigma_{xy}$ at different sample sizes can be scaled
onto a scaling curve if plotted as a function of $L/\xi$ at fixed disorder 
strength. Here $\xi$ is the thermodynamic localization length,
$\xi=(\epsilon_0/|\epsilon-\epsilon_c|)^{\alpha}$ with $\alpha\approx 
2.3$. But it was expected that the shape of the scaling curve may, more or less,
depend on disorder potential. By contrast, we find that the scaling 
of conductances here is actually universal and the data at different $W$'s
all collapse onto a universal curve if we choose a right energy scale
$\epsilon_0$ at different $W$'s. For example, as shown in Fig. 3a,
 $\sigma_{xx}$ at $W=1$ and $W=4$ for  $L=128$ and $L=160$ can be scaled 
onto the same scaling function of $L^{1/\alpha} \Delta \epsilon/\epsilon_0$ by choosing $\epsilon_0 (W=4)=3.57$ with $\epsilon_0 (W=1)$ chosen as the unit. 
Notice that the data shown in Fig. 3a are symmetric
about its critical point $\epsilon_c$ as shown in the insert.
Similar behavior is also found for $\sigma_{xy}$.
In addition to the universal scaling of $\sigma_{xx}$ and $\sigma_{xy}$,
we also found that the scaling curve of $\sigma_{xx}^2$ for sample sizes 
from $L=24,32,64,128$ to $160$ coincides with the scaling curve of 
$\sigma_{xy}*(1-\sigma_{xy})$ in Fig. 3b,
independent of the disorder strength $W$. [Note that while the maximum width for
$\sigma_{xx}$ calculation reaches $L=160$, the largest sample width attainable for $\sigma_{xy}$ is much smaller ($L=48$) in the present work and in 
literature\cite{bhatt}).] This 
scaling relation guarantees the semi-circle law (\ref{3}) and the exact 
quantization of $\rho_{xy}$ in the whole $1 \rightarrow 0$ transition region. 
Thus we find that both the exact quantization of
$\rho_{xy}$ and the reflection symmetry of $\rho_{xx}$ are 
the consequences of a universal scaling satisfied by the conductances 
which is independent of the details of the underlying model like the 
particle-hole symmetry and the
lattice effect (we have checked weaker magnetic fields with $M=16$ and $24$ and
found essentially consistent results).  

It is noted that in the above numerical calculations the resistances and 
conductances are shown to be scaling functions with regard to the 
scaling variable $\xi/L$ around the critical point. Here the zero-temperature 
transition is driven by Fermi energy with 
$\xi=(\epsilon_0/|\epsilon -\epsilon_c|)^{\alpha}$. On the 
other hand, experiments have been done at finite temperature where the thermodynamic sample size is cut-off by a finite length scale $L_{in}$ representing the de-phasing effect such that the scaling variable 
becomes $\xi/L_{in}$. When magnetic field instead of the Fermi energy is tuned
in experiment, one has $\xi=\xi_0|B-B_c|^{-\alpha}\propto |n_{\nu}-n_{\nu c}|^{-
\alpha}$. Thus, by using such a scaling variable the reflection symmetry found 
in our numerical calculations is related to the observed one in experiments as
given in (\ref{1}).

We have also checked the behavior of $\rho_{xy}$ vs. $\rho_{xx}$ in whole
regime. As shown in Fig. 4, the quantized $\rho_{xy}$ persists from the
QH plateau regime with $\rho_{xx}<1$ to the transition regime with $\rho_{xx}$ 
as large as $5$ which is comparable to the experimental data.\cite{shar1,shar2}
Deep into the insulating phase with $\rho_{xx}\rightarrow \infty$, $\rho_{xy}$ becomes divergent too. The latter behavior is consistent with the result
obtained by the network model calculation with including quantum 
interference.\cite{prya2} 

Finally, we have considered a case in which the reflection symmetry of 
$\rho_{xx}$ and
the quantization of $\rho_{xy}$ near the $\nu=1\rightarrow 0$ transition regime
disappear. It has been previously shown that in the tight-binding lattice model
the higher QH plateaus can be destroyed first at strong disorders while the 
lowest plateau remains robust\cite{dns}. In this case, there exists
a second $\nu=1\rightarrow 0$ transition between the QH state to a high filling
insulator at a critical disorder $W_c$ for a given Fermi energy. 
At $W < W_c$, $\rho_{xy} $ stays quantized in the $\nu=1$ plateau regime. But 
beyond $W_c$, $\rho_{xy}$ starts to deviate from the quantization and 
continuously drop from the quantized value approaching to the classical value
$\rho_{xy}=h/e^2 n_{\nu}$, while the reflection symmetry of $\rho_{xx}$ 
is no longer present. Thus, the well-defined 
Hall plateau(s) above the Fermi energy may be a crucial condition for
both $\rho _{xy} $ quantization and the reflection symmetry of $\rho_{xx}$ to 
occur. This second transition has been related\cite{dns} to the experimentally 
observed $\nu=1\rightarrow 0$ transition at weak magnetic field side, and detailed properties will be discussed elsewhere. 

We conclude that the reflection symmetry of resistance and the quantization of
Hall resistance are intrinsically present in the $\nu=1\rightarrow 0$
transition in a two-dimensional non-interacting electron system at 
strong magnetic field. Our calculations suggest that they are determined by
a universal scaling of conductances which is independent 
of the particle-hole symmetry in the LLL. The present numerical results explain
the recent experimental measurements and provide very useful insight for 
further theoretical studies.  
  
{\bf Acknowledgments} - The authors would like to thank L. P. Pryadko,
S. Subramanian, C. S. Ting, Z. Q. Wang, X. G. Wen, X. C. Xie, and K. Yang
for stimulating and helpful discussions. We also would like to thank the hospitality of the National Center for Theoretical Science in Taiwan where the 
present work was partially completed. The present work is supported by the State 
of Texas through ARP grant No. 3652707 and Texas Center for Superconductivity at University of Houston.

Fig. 1 The quantization of the Hall resistivity near the $\nu=1\rightarrow 0$
transition: $\rho_{xy}$ versus $\sigma_{xy}$ at sample width $L=8$ ($*$); $16$ ($+$); $24$ ($\bullet$) and $32$ ($\diamond$) at disorder strength $W=1$ (a) and
$W=4$ (b). The inserts show the density of states $D_L(\epsilon)$; $\epsilon_c$
denotes the critical point. 

Fig. 2 The reflection symmetry of $\rho_{xx}$: $\rho_{xx}(\Delta \epsilon)$ 
and $1/\rho_{xx}(-\Delta\epsilon)$ coincide at $W=1$ and $W=4$ with sample 
width $L=32$ ($\Delta\epsilon=\epsilon-\epsilon_c$).
 
Fig. 3 (a) $\sigma _{xx}$ as a scaling function of $L^{1/\alpha}\Delta 
\epsilon/\epsilon_0$. Here 
$\alpha=2.3$ and $\epsilon_0=3.57$ at $W=4$ ($\epsilon_0$ at $W=1$
is chosen as the unit). The insert shows that $\sigma_{xx} $ is a symmetric 
function of $\Delta \epsilon $ around the origin ($L=128$ and $W=4$).
(b) $\sigma_{xx}^2$ and $\sigma_{xy}*(1-\sigma_{xy})$ collapse onto a scaling
function of $L^{1/\alpha}\Delta \epsilon/ \epsilon_0 $ with 
$L=24,32,64,128$, and $160$, and $W=1$, $4$.

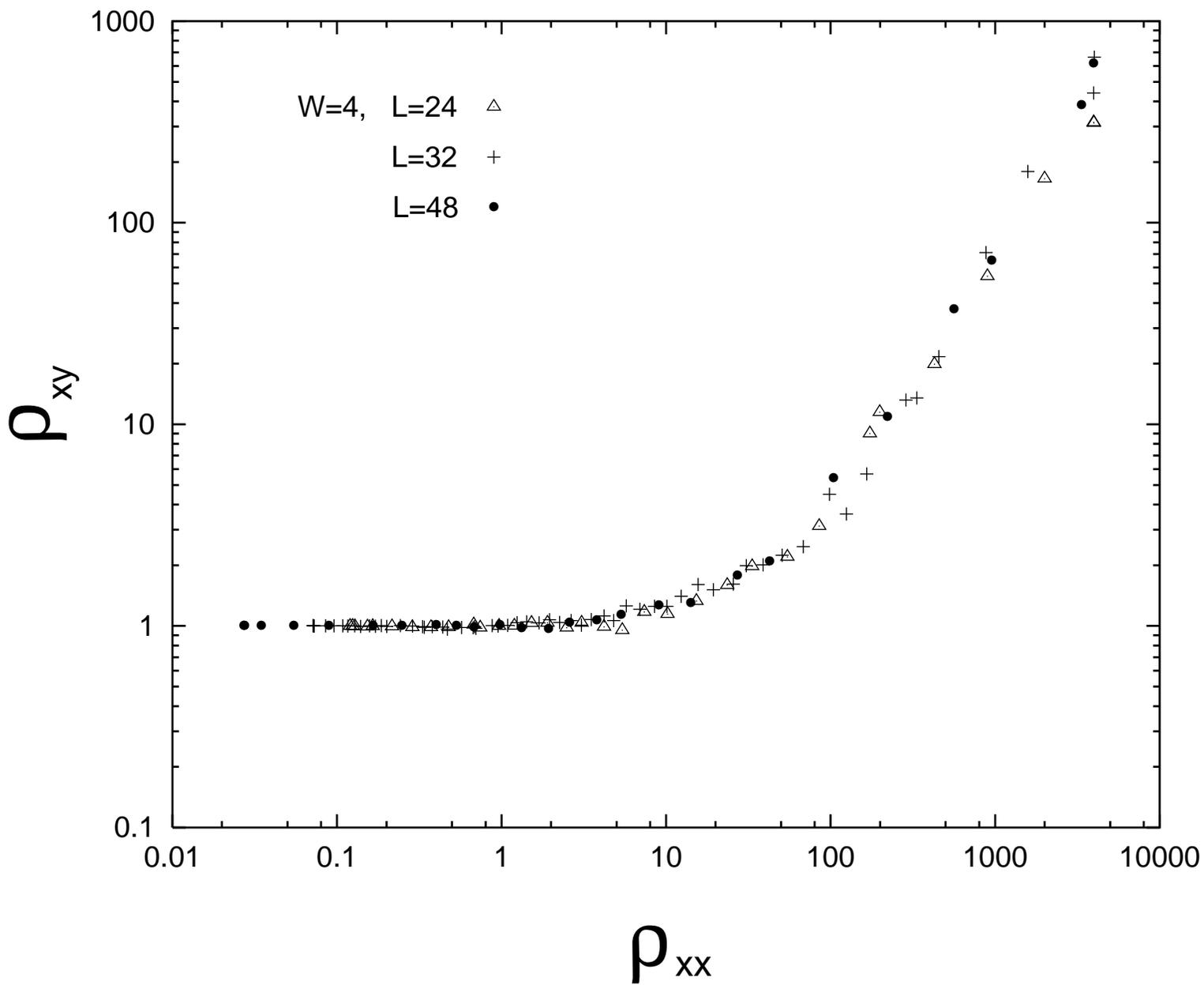
Fig. 4. $\rho_{xy}$ vs. $\rho_{xx}$ in whole regime at fixed $W=4$. 

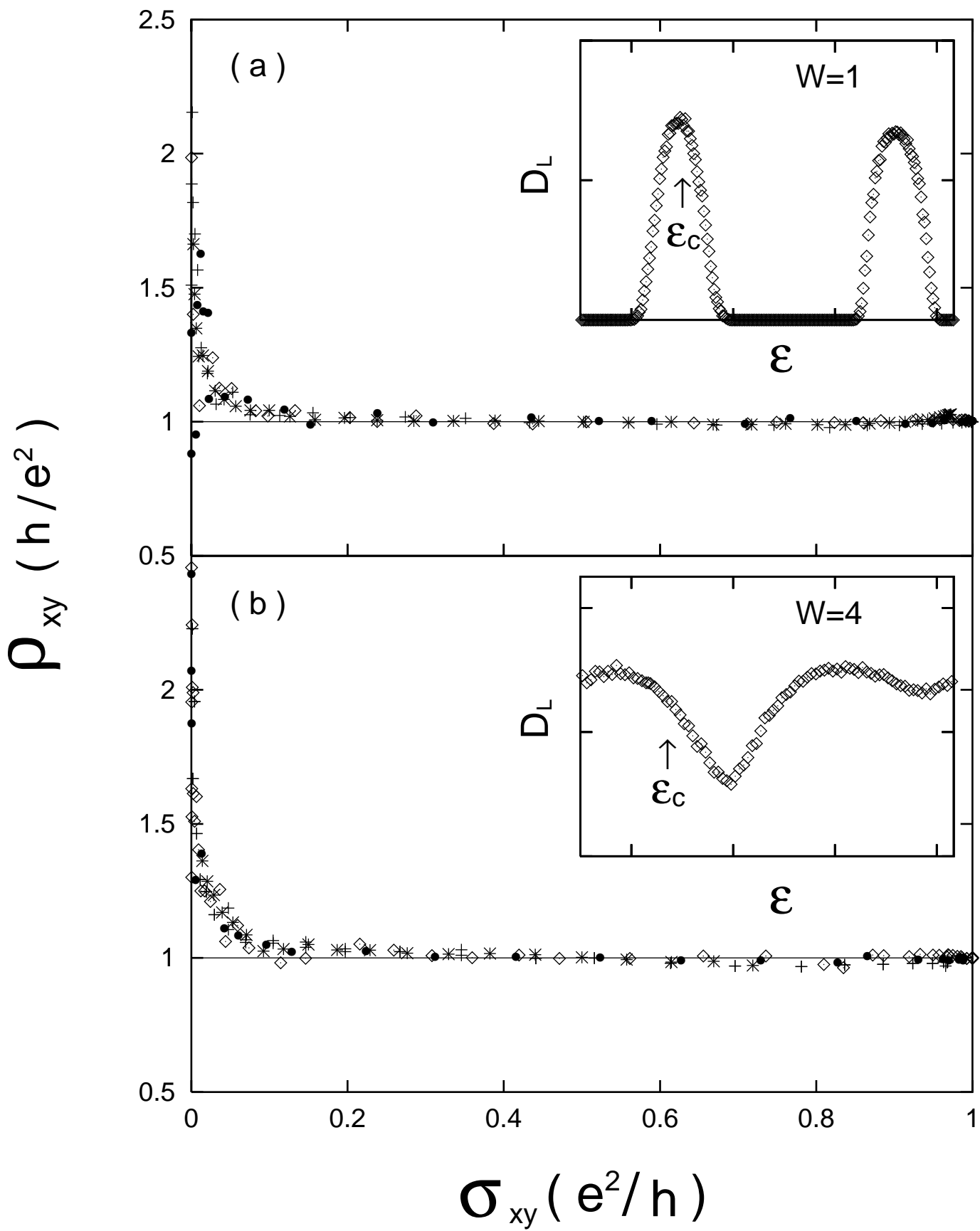
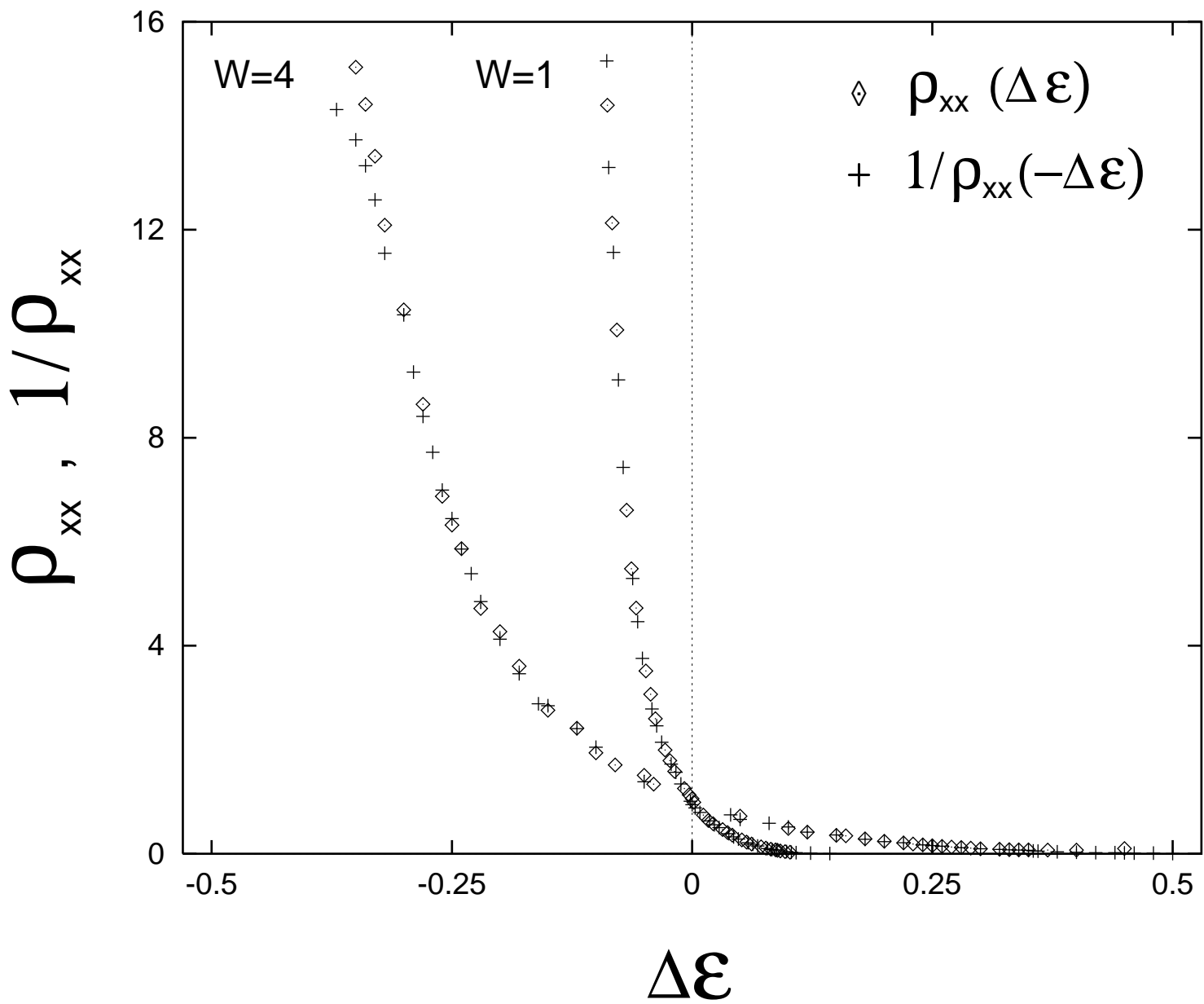
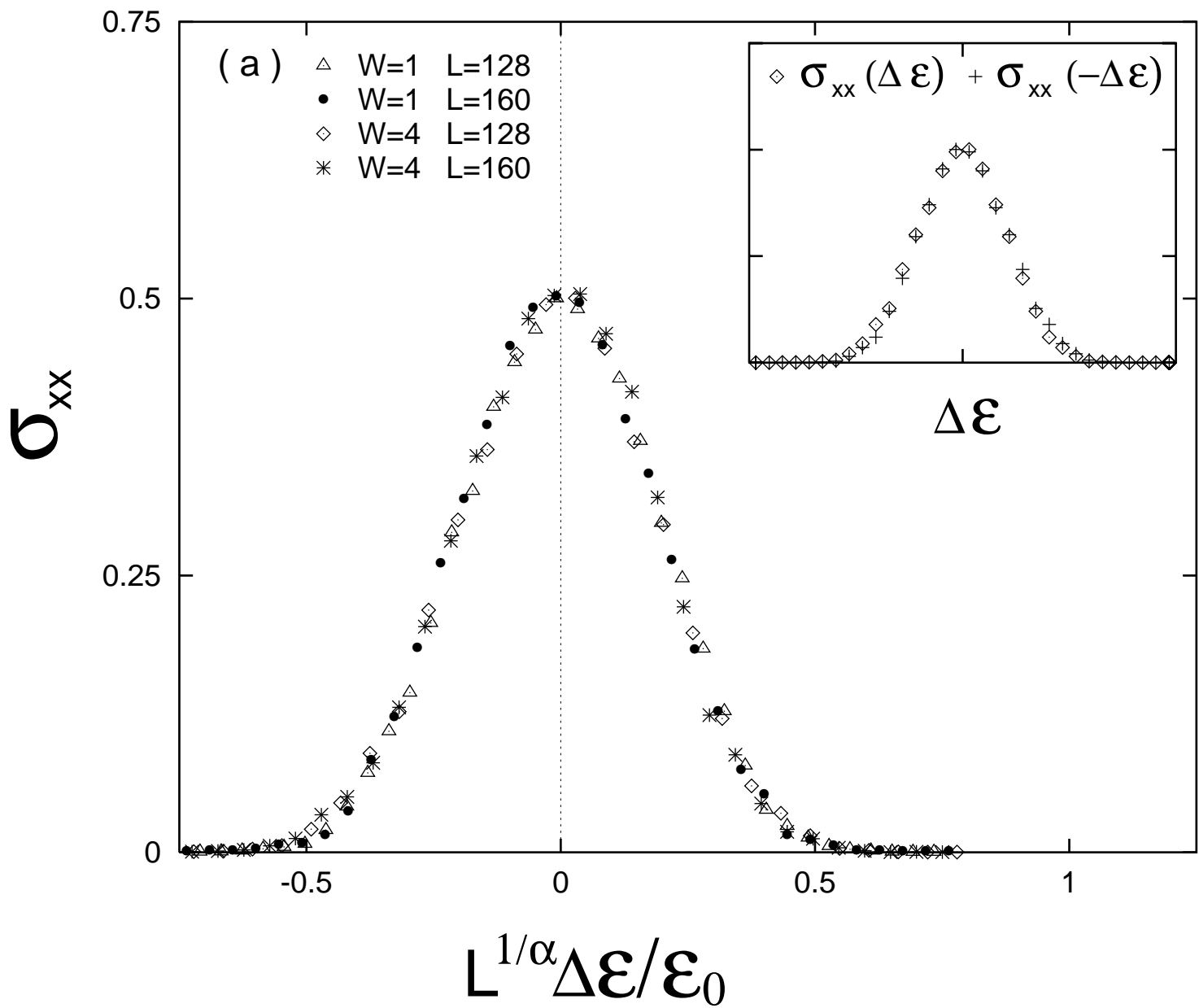
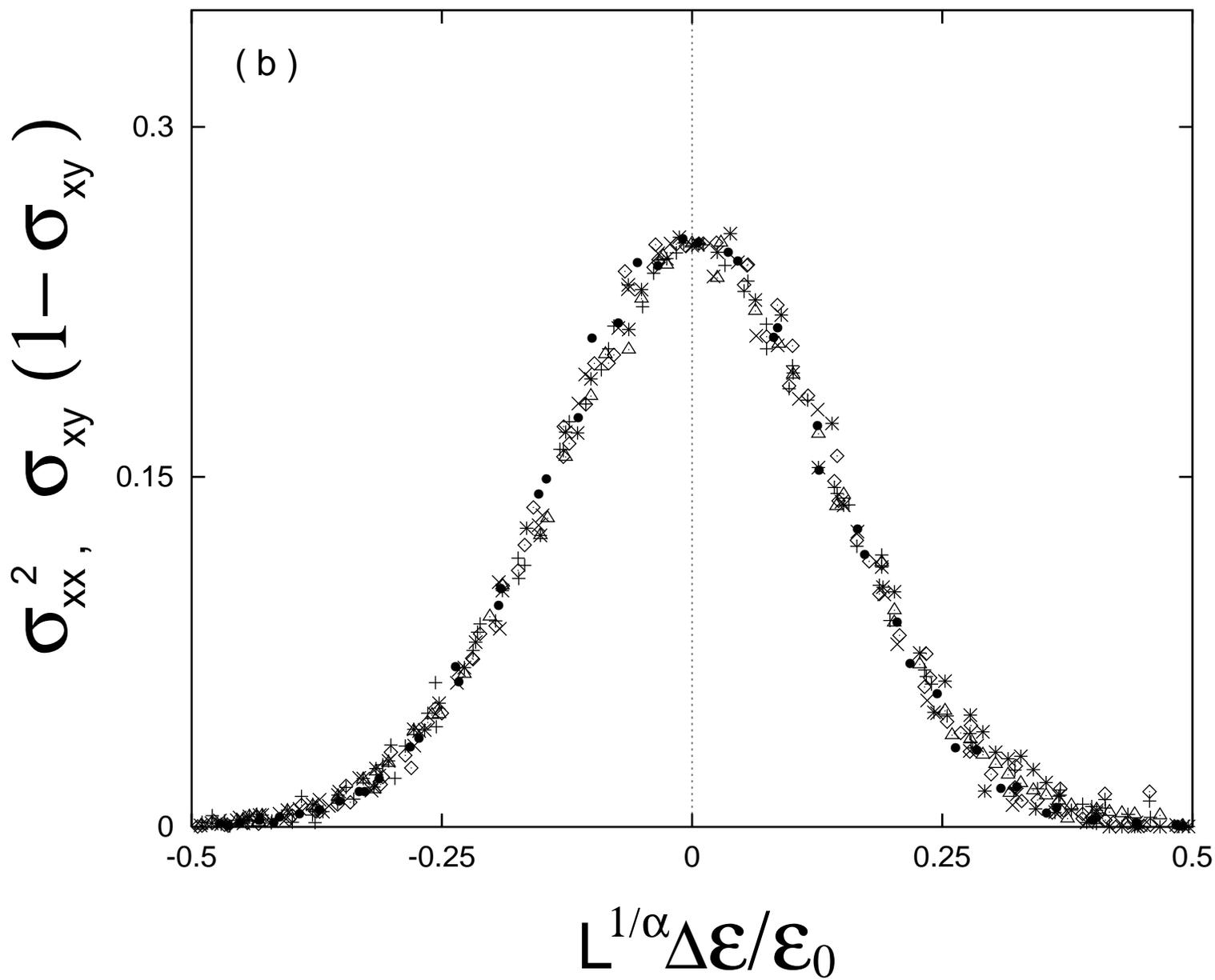
\end{document}